\documentclass[prd,aps,showpacs,tightenlines,superscriptaddress,amsmath,amssymb,amsfonts,twocolumn]{revtex4-2}
\usepackage{amssymb,amsmath,amsthm,amsbsy,epsfig,color,graphicx,times}
\usepackage[ansinew]{inputenc}
\usepackage[english]{babel}
\usepackage{color}
\usepackage{braket}
\usepackage{tabularx}
\usepackage{array}
\usepackage{slashed}

\begin{document}

\title{The impact of the $X_{17}$ boson on particle physics anomalies: muon anomalous magnetic moment, Lamb shift, $W$ mass and dark charges}

\author{A. Capolupo}
\email{capolupo@sa.infn.it}
\affiliation{Dipartimento di Fisica ``E.R. Caianiello'' Universit\`{a} di Salerno, and INFN --- Gruppo Collegato di Salerno, Via Giovanni Paolo II, 132, 84084 Fisciano (SA), Italy}

\author{A. Quaranta}
\email{anquaranta@unisa.it}
\affiliation{Dipartimento di Fisica ``E.R. Caianiello'' Universit\`{a} di Salerno, and INFN --- Gruppo Collegato di Salerno, Via Giovanni Paolo II, 132, 84084 Fisciano (SA), Italy}

\author{R. Serao}
\email{rserao@unisa.it}
\affiliation{Dipartimento di Fisica ``E.R. Caianiello'' Universit\`{a} di Salerno, and INFN --- Gruppo Collegato di Salerno, Via Giovanni Paolo II, 132, 84084 Fisciano (SA), Italy}

\begin{abstract}
We show that the $X_{17}$ vector boson, introduced to explain the $^8Be$ anomalous decay, could play a crucial role in the explanation of the muon's (electron's) anomalous magnetic moment and the muonic Lamb shift.
We further constrain the possible kinetic mixing with the $U(1)_Y$ boson of the Standard Model by using the latest avaible data on the $W$ boson mass.
\end{abstract}

\maketitle

Physics beyond the Standard Model (SM) of particles has undergone significant advancements in recent years. This pursuit aims to elucidate several phenomena including particle mixing and oscillations, matter-antimatter asymmetry, the nature of dark matter and dark energy \cite{BSM1,BSM2,BSM3,BSM4,BSM5,BSM6, BSM7, BSM8, BSM9, BSM10, BSM11, BSM12, BSM13, BSM14, BSM15, BSM16, BSM17, BSM18, BSM19, BSM20, BSM21, BSM22, BSM23, BSM24}.
Another phenomenon whose explanation presumably requires physics beyond the Standard Model is the muon magnetic moment anomaly \cite{Marciano2016, Cazzaniga2021, NL}.
We denote the anomalous magnetic moment $a_l$ for the lepton $l=e,\mu,\tau$,  corresponding to the difference between the quantum field theoretic prediction and the Dirac value, i. e. $a_l = \frac{g_l-2}{2}$.
The disagreement between the theoretical SM calculations of $a_l$ and the experimental measurements $a_l^{EXP}$  could indicate contributions from physics beyond the SM.
Considering the electron magnetic moment, the experimental value $a^{EXP}_e=1159652180.73\times 10^{-12}$ is precise to $0.24$ parts-per-billion \cite{Hannake} and agrees with the theoretical prediction to an astounding precision of   $10^{-13}$ \cite{Nature2020}: $\Delta a_e = a_{e, EXP} - a_{e, SM} = (4.8  \pm 3.0) \times 10^{-13}$.
On the other hand, the Standard Model prediction for the muon magnetic moment is much less in agreement with the experimental data. Indeed, recently, the Muon g-2 collaboration's latest experimental results \cite{Abi2021} have confirmed a discrepancy for $a_\mu$ with respect to the Standard Model value \cite{Patrignani2016,Aoyama2020}. The combined data from the Brookhaven \cite{Bennett2004} and Fermilab Muon g-2 \cite{Abi2021} experiments indicate a $4.2 \sigma$ discrepancy: $\Delta a_{\mu} = a_{\mu, EXP} - a_{\mu, SM} = (251 \pm 59) \times 10^{-11}$.

Also other precision tests of quantum electrodynamics (QED) reveal tensions with the Standard Model predictions. The $2S_{1/2}-2P_{1/2}$ Lamb shift for muonic atoms \cite{muon1, muon2, muon3, muon4}, for instance, displays a significant deviation from the Standard Model. In particular, the Lamb shift between $2S$ and $2P$ levels for muonic hydrogen and muonic deuterium differ from the Standard Model expected values by: $\delta E_\mu^H=(-0.363,-0.251)\ \mathrm{meV}$ \cite{muon1, muon2} and  $\delta E_\mu^D=(-0.475,-0.337) \ \mathrm{meV}$ \cite{muon3, muon4}. This in turn has led to the so-called proton radius puzzle \cite{PRP1, PRP2, PRP3, PRP4}, where a significantly smaller proton radius was observed compared to previous measurements in regular hydrogen. This discrepancy points to potential gaps in our understanding of hadronic structure or hints at new physics beyond the Standard Model.
Additional hints of new physics may also come from the gauge bosons sector.
Electroweak analysis, including $Z-$pole data and measurements of $m_{top}$ and $m_h$ provides a well-determined mass of $m_{W,SM}=80361\pm 6 \ \mathrm{MeV}$ for the $W$ boson that mediates eletroweak interactions. Therefore, precise measurement of the $W$ boson mass is crucial for testing the internal consistency of the Standard Model.

Recently, the CMS collaboration reported an accurate measurement of the $W$ mass, which is $80360 \pm 9.9 \ \mathrm{MeV}$ \cite{W1}. The $W$ boson mass, at tree level is equal to $g\nu/2$, where $\nu=246\ \mathrm{GeV}$ is the vacuum expectation value of the Higgs field and $g$ denotes the weak isospin coupling parameter. If new particles are present, the $W$ boson mass is expected to receive additional loop corrections.
As we will show, the QED anomalies described above as well as tensions regarding the $W$-boson mass may be explained in terms of a new light boson. A significant impetus for the search for a new 'fifth' force carrier weakly coupled to SM particles was provided by the recent observation of a $\sim 7\sigma$ excess of events in the angular distribution of $e^+e^-$ pairs produced during the nuclear transition of the excited $^8Be^*$ to its ground state via internal $e^+e^-$ pair creation\cite{X1,X2}. The latest results from the ATOMKI group report a similar excess at approximately the same invariant mass in the nuclear transitions of another nucleus, $^4He$. It has been speculated that this anomaly can be interpreted as the emission of a protophobic gauge boson corresponding to a new $U(1)_X$ symmetry, the $X_{17}$ boson, so christened due to its mass being around $17 \ \mathrm{MeV}$ and which decays into $e^+e^-$ \cite{X3,X4}.
Several experiments and theoretical analyses have been performed to assess the viability of this new light boson \cite{X5,X6,X7,X8,X9,X10,X11,X12,X13}.

In the present work, we demonstrate that the presence of the $X_{17}$ could, in principle, explain the muon $g-2$ anomaly and the Lamb shift anomalies. We moreover consider the kinetic mixing of the $X_{17}$ to the hypercharge $U(1)_Y$ boson of the standard Model, determining the corresponding $W$ mass correction, and establishing an upper bound on the kinetic mixing parameter from the latest measurement of the $W$ boson mass.
Specifically, we compute the one loop correction to the magnetic dipole moment induced by the presence of $X_{17}$ boson and derive an upper bound on the coupling constant of the $X_{17}$ to electron and muon. Moreover, using the thus determined coupling for the muon and considering the nonrelativistic potential induced by the $X_{17}$ for muonic atoms, we derive an upper limit on the coupling costants of the $X_{17}$ to proton and neutron. Finally, taking into account a possible kinetic mixing between $X_{17}$ and the hypercharge $U(1)_Y$ boson, we deduce the contribution of the former to the $W$ boson mass.
In line with the latest experimental data we constrain the kinetic mixing parameter.

\begin{figure}[t]
\label{fig:fey}
\includegraphics[width=\columnwidth]{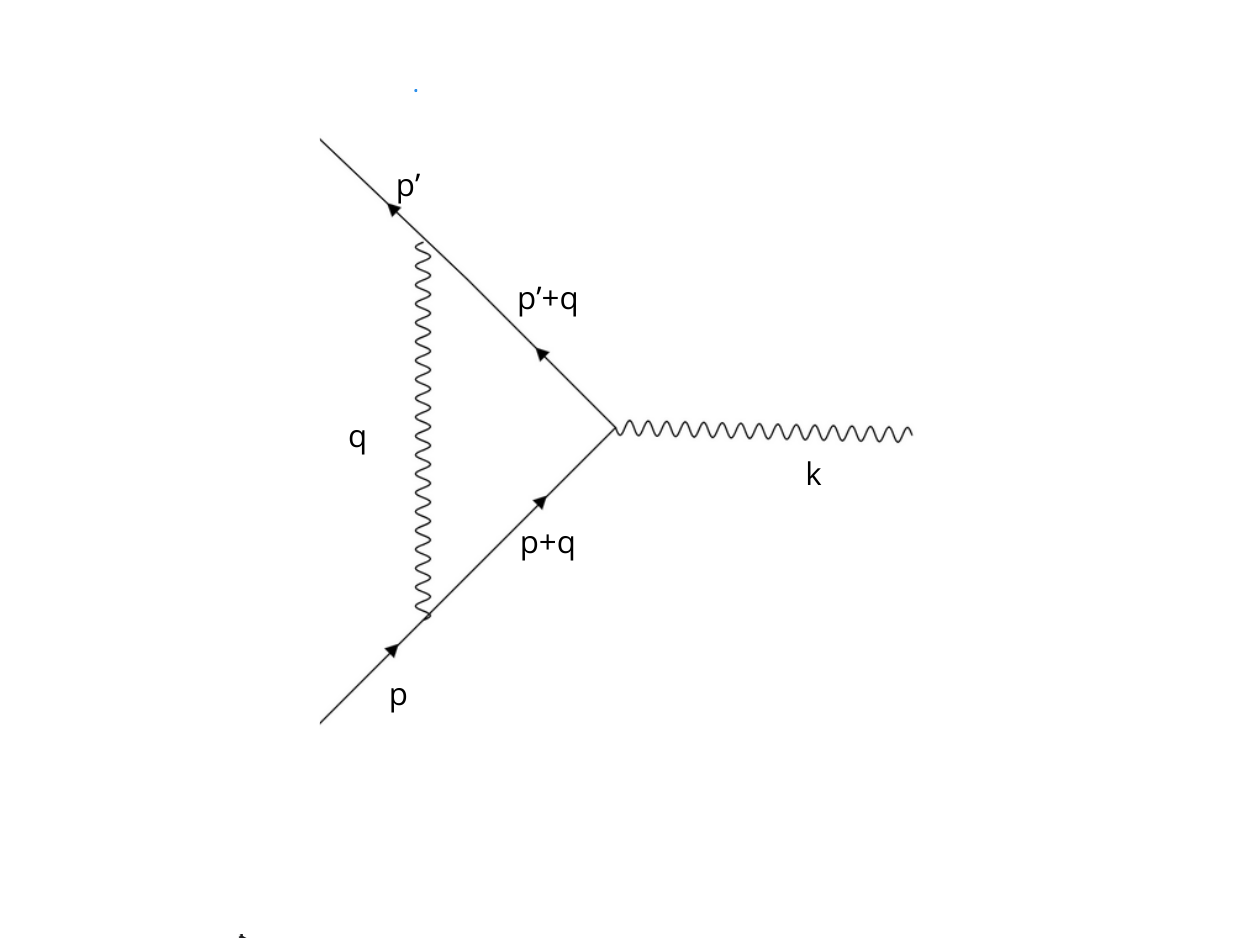}
\caption{Reference Feynman diagram for the calculation of the one loop correction to the magnetic dipole moment.}
\end{figure}

\vspace{2mm}

{\it Lepton magnetic moment} -- In presence of a new vector boson, like $X_{17}$, coupled to charged leptons, there emerges a correction to the $g$-factor.
The relevant diagram at one loop is that of Fig 1 and the vertex correction is
\begin{eqnarray}\label{Vertex}
\nonumber &&\delta \Gamma^{\mu} (q) = \int \frac{d^4 q}{(2\pi)^4} \bar{u} (p') (-i \epsilon_l  \gamma^{\lambda})\frac{i(\slashed{p}'+\slashed{q}+m_l)}{(p'+q)^2-m_l^2}(e \gamma^\mu) \\ \nonumber
&&\times \frac{i(\slashed{p}+\slashed{q}+m_l)}{(p+q)^2-m_l^2} (-i \epsilon_l  \gamma^\nu)\frac{i}{q^2+M^2_X}\biggl(-g^{\nu\lambda}+\frac{q^\nu q^\lambda}{M^2_X}\biggr) u(p) ,   \\
\end{eqnarray}
where $\epsilon_l$ is the coupling constant  between the $X_{17}$ vector boson and lepton $l$.
The resulting correction to the $g$-factor is analogous to that of a dark photon \cite{Posp} and is
\begin{equation} \label{F2}
    a_l^X=\frac{\alpha}{2\pi} (\epsilon_l m_l)^2 \int_0^1 dx \frac{x^2(1-x)}{m_l^2x^2+M^2_X(1-x)} .
\end{equation}
 This equation can be written in a simpler fashion, introducing the adimensional parameter $\lambda_l=m_l^2/M^2_X$:
 \begin{equation}
 \label{deltaa}
    a_l^X= \frac{\alpha}{2\pi} \epsilon_l^2 \int_0^1 dx \frac{x^2(1-x)}{a^2x^2-x+1}=\frac{\alpha}{2\pi} \epsilon_l^2 f(\lambda_l).
 \end{equation}
 Therefore, Eq.\eqref{deltaa} can be understood as an effective shift of the electrodynamic coupling costant by:
 \begin{equation}
     \Delta\alpha=2\pi a_l^X .
 \end{equation}
Since the correction explicitly depends on the ratio between the mass of the lepton and that of the boson, it is evident that the correction becomes more significant as the lepton's mass increases. Thus, this correction is almost negligible for the electron, becomes noticeable for the muon, and can be expected to be predominant when considering the tau.
The $X_{17}$ boson correction might then explain, at least partially, the  muon $(g-2)$ anomaly.
Attributing the difference in the observed values of $a_l$ and the standard model computation to the $X_{17}$ correction, we now determine the upper bounds on $\epsilon_l$ for $l= e,\mu$.
The relevant inequality we will use is
\begin{eqnarray}\label{momm}
 \delta a_\mu= a_{\mu , EXP} - a_{\mu, SM} \leq 2.51\times 10^{-9} \ ,
\end{eqnarray}
in which the latest available average values for the Standard Model prediction \cite{Patrignani2016} for $a_{\mu, SM}$ and the experimental result \cite{Abi2021} for $a_{\mu , EXP}$ are considered.
Notice that there are two possibilities: in principle the dark charges may be flavor-blind or flavor dependent.
However it can be shown that \cite{Feng} flavor-blind couplings would lead to unobserved charged lepton oscillations.
Therefore we consider the case of flavor-dependent couplings.
 Considering a mass of the $X_{17}$ of $17\  \mathrm{MeV}$, the upper bound on the coupling of the $X_{17}$ to the muon, as resulting from the comparison between the inequality \eqref{momm} and Eq. \eqref{deltaa} is $|\epsilon_\mu| <2.154 \times 10^{-4}$. Notice that, since the correction of Eq. \eqref{deltaa} is quadratic in $\epsilon_l$, we cannot establish the absolute sign of $\epsilon_{\mu}$ from this analysis. 
 
The coupling to the $X_{17}$ boson can likewise explain the electron magnetic moment anomaly. The relevant inequality is:
\begin{eqnarray}\label{mome}
 \delta a_e=a_{e , EXP} - a_{e, SM} \leq 4.8 \times 10^{-13},
\end{eqnarray}
where we have employed the direct experimental measurement $a_{e,EXP}$ \cite{Nature2020}.
In this case, the upper bound on the coupling of the $X_{17}$ to the electron is $|\epsilon_e|< 1.02 \times 10^{-4}$.
We note that these results are in agreement with the recent analyses performed regarding neutrino-$X_{17}$ interactions as shown in \cite{exp1}.

\vspace{2mm}

{\it Lamb shift} -- The most recent measurements of the Lamb shift in muonic atoms show a deviation with respect to the predictions of the Standard Model. In particular, for muonic hydrogen, the difference between the experimental results and the Standard Model prediction is \cite{muon1, muon2}:
$\delta E_\mu^{H}=(-0.363, -0.251) \  \mathrm{meV}$
and for the muonic deuterium \cite{muon3, muon4}:
$\delta E_\mu^D=(-0.475, -0.337)  \ \mathrm{meV}$.

We now explore the possibility that the tension between experimental measurements and theoretical predictions can be solved by means of the interaction with the $X_{17}$ boson.
Since  $X_{17}$ is a vector boson, with vectorlike couplings, the nonrelativistic potential between the proton and muon due to the $X_{17}$ exchange is:
\begin{equation}
\label{Pot}
    V_X(r)=\frac{\epsilon_\mu \epsilon_p}{e^2} \frac{\alpha e^{-M_X r}}{r},
\end{equation}
where $\epsilon_\mu$ can be estimated as shown above from the analysis of $g-2$ anomaly and $\epsilon_p$ is the coupling of the $X_{17}$ to the proton.
This potential gives an additional contribution to the Lamb shift in the $2S_{1/2}-2P_{3/2}$ transition, which, using first order perturbation theory, is given by:
\begin{equation}
\label{deltaH}
\begin{split}
    \delta E^H_X&=\int dr r^2V_X(r)(\|R_{20}(r)\|^2-\|R_{21}\|^2)\\
    &= \frac{\alpha}{2 a_H^3} \biggl(\frac{\epsilon_\mu \epsilon_p}{e^2}\biggr) \frac{f(a_H M_X)}{M_X^2} \ .
    \end{split}
\end{equation}
Here $f(x)=\frac{x^4}{(1+x)^4}$ and $a_{H}=(\alpha m_{\mu p})^{-1}$ is the Bohr radius of the system, $m_{\mu H}$ is the reduced mass of the muonic hydrogen and $R_{nl}$ are the radial wave function. Fixing the value of $\epsilon_{\mu}$ equal to the upper bound found from the previous analysis on the muon $g-2$ anomaly, and inserting the experimental value for $\delta E^{H}_{\mu}$, we can invert Eq. \eqref{deltaH} to deduce $\epsilon_p$. This yields an upper bound on the (modulus of the) coupling to the proton in terms of the $X_{17}$ mass $M_X$ and of the Lamb shift deviation $\delta E^{H}_{\mu}$, as depicted in Fig. 2. Since $\epsilon_p < 0$ for $\epsilon_{\mu} > 0$, the contour plot of Fig. 2 displays the lower bound on $\epsilon_p$, as a function of  the mass of $X_{17}$ and experimental data on muonic hydrogen  \cite{muon1, muon2}. The values are deduced by inverting Eq. \eqref{deltaH} in correspondence of $\epsilon_{\mu} = 2.154 \times 10^{-4}$, as derived from the analysis on the $g-2$ anomaly.  Considering a range of values for $M_{X}$ between $[16.7, 17.2] \ \mathrm{MeV}$, and for $\delta E_\mu^H$ between $[-0.363, -0.251]\ \mathrm{meV}$, the range of values for the lower bound on the coupling $\epsilon_p$ will be between $[-0.04260,-0.02982]$. Consequently the upper bound on $|\epsilon_p|$ ranges in $[0.02982,0.04260]$.

As remarked above, there is a sign ambiguity on $\epsilon_{\mu}$: the values of Fig. 2 are obtained by assuming a positive muon coupling $\epsilon_{\mu} > 0$, which leads to a negative proton coupling $\epsilon_{p} < 0$. The general conclusion is that $\epsilon_p$ and $\epsilon_{\mu}$ must have opposite sign in order to account for the Lamb shift anomaly in muonic hydrogen. As clear from Eq. \eqref{deltaH} and from Fig. 2, the lower bound on $\epsilon_p$ is linear in the Lamb shift deviation $\delta E_\mu^H$ and approximately quadratic in the boson mass $M_X$.

Using the analogous of Eq.\eqref{deltaH} for muonic deuterium, we derive the following additional contribution to the
Lamb shift:
\begin{equation}\label{DeltaD}
    \delta E^D_X= \frac{\alpha}{2 a_D^3} \biggl(\frac{\epsilon_\mu (\epsilon_p+\epsilon_n)}{e^2}\biggr) \frac{f(a_D M_X)}{M_X^2}.
\end{equation}
Here $a_{D}=(\alpha m_{\mu D})^{-1}$ is the Bohr radius of the system, and $m_{\mu D}$ is the reduced mass of the muonic deuterium. Inserting the values of $\epsilon_{\mu}$ as derived from the $g-2$ analysis and of $\epsilon_p$ as estimated from the Lamb shift on muonic hydrogen, we can invert Eq. \eqref{DeltaD} to obtain an estimate on the upper bound on $\epsilon_{n}$.

In Fig. 3, we report a contour plot of the upper (lower) bound on the coupling constant $\epsilon_n$, depending on its sign $\epsilon_n > 0$ ($\epsilon_n < 0$) as a function of experimental data on muonic hydrogen \cite{muon1, muon2} and muonic deuterium \cite{muon3, muon4}, as deduced from equation \eqref{DeltaD}, and in correspondence with $M_X = 17 \ \mathrm{MeV}$. Considering a range of values for for $\delta E_\mu^H$ between $[-0.363, -0.251]\ \mathrm{meV}$, and for $\delta E_\mu^D$ between $[-0.475, -0.337]\ \mathrm{meV}$, the range of values for the coupling $\epsilon_n$ will be between $[-0.010, 0.010]$. Overall $|\epsilon_n| \leq 10^{-2}$.

Once again there is an overall sign ambiguity due to the unresolved $\epsilon_{\mu}$ sign.
Interestingly, in order to account for both the discrepancies $\delta E_\mu^H$ and $\delta E_\mu^D$, depending on the range considered, $\epsilon_p$ and $\epsilon_n$ might have either the same or the opposite sign. 
The upper (lower) bound on $\epsilon_n$ is clearly linear in both Lamb shift deviations $\delta E_\mu^H, \delta E_\mu^D$ as evident from Eqs. \eqref{deltaH} and \eqref{DeltaD}. 

\begin{figure}[t]
\includegraphics[width=\columnwidth]{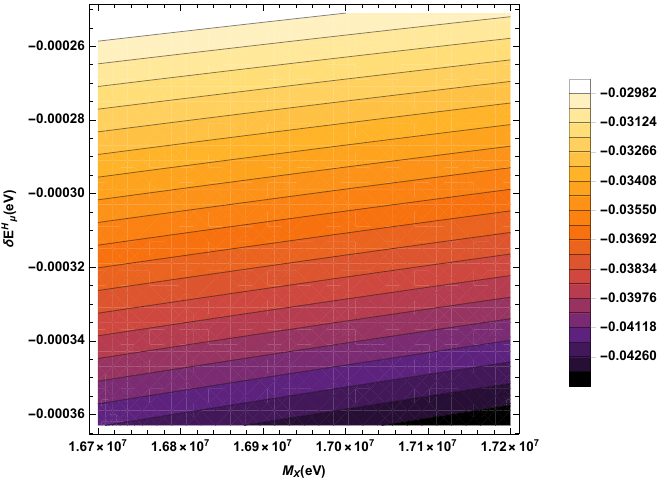}
\caption{ (Color online) Contour plot of the lower bound for the coupling constant $\epsilon_p$ of $X_{17}$ to the proton as a function of the mass of $X_{17}$ and the experimental data on muonic hydrogen Lamb shift  $\delta E_\mu^H$ \cite{muon1, muon2}, as obtained from equation \eqref{deltaH}. The legend on the side shows the values of the coupling constant $\epsilon_p$ for different values of $M_X$ and $\delta E_\mu^H$. We consider $\epsilon_{\mu} \simeq 2.154 \times 10^{-4}$ .}. 
\end{figure}

\begin{figure}[t]
\includegraphics[width=\columnwidth]{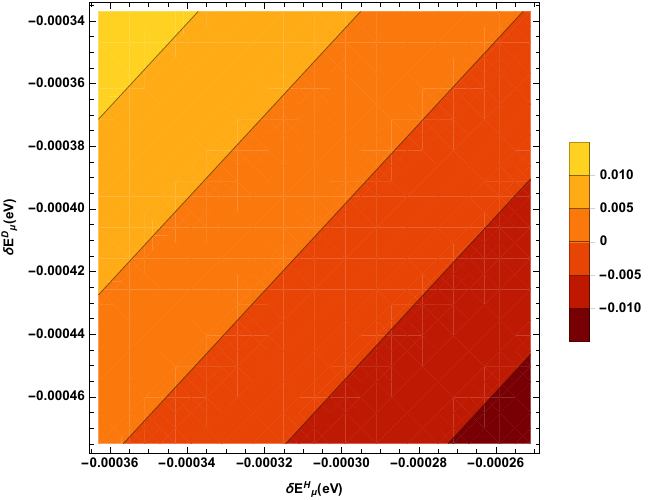}
\caption{ (Color online) Contour plot of the upper (lower) bound for the coupling constant $\epsilon_n > 0$ ($\epsilon_n < 0$) of $X_{17}$ to the neutron in terms  experimental data on  muonic hydrogen shift $\delta E_\mu^H$  \cite{muon1, muon2} and muonic deuterium $\delta E_\mu^D$ \cite{muon3, muon4}, as deduced from equation \eqref{DeltaD}. The legend on the side shows the values of the coupling $\epsilon_n$ for different values of $\delta E_\mu^H$ and $\delta E_\mu^D$. We have used $|\epsilon_{\mu}| \simeq 2.154 \times 10^{-4}$ and we have assumed $\epsilon_\mu>0$. The coupling to the proton $\epsilon_p$ is fixed according to Eq. \eqref{deltaH}, ranging in the values shown in Fig. 2. The boson mass is here fixed to the value $M_X = 17 \ \mathrm{MeV}$.}
\end{figure}

\vspace{2mm}

{\it W mass} -- For this analysis we follow the method used in \cite{W00} for the kinetic mixing between the hypercharge boson $B$ and the dark photon, suitably modified for the $X_{17}$.
By using the latest data on the $W$ boson mass \cite{W1}, which leave an uncertainty of $9.9\ \mathrm{MeV}$, we constrain possible kinetic mixing. Indeed, if $X_{17}$ behaves as a dark photon or a dark $Z$, it can experience mixing with neutral bosons, and in particular with the hypercharge boson $B_{\mu}$.
The relevant part of the Lagrangian is:
\begin{equation}
\mathcal{L}\supset \frac{1}{4} \hat{B}_{\mu\nu}\hat{B}^{\mu\nu}-\frac{1}{4}\hat{X}_{\mu\nu}\hat{X}^{\mu\nu}+\frac{\xi}{2c_W}\hat{X}_{\mu\nu} \hat{B}^{\mu\nu} +\frac{1}{2}M^2_{X,0}\hat{X}_{\mu}\hat{X}^\mu ,
\end{equation}
where $X_{\mu}$ is the $X_{17}$ field, $c_W=\cos{\theta_W}$, $\theta_W$ is the Weinberg angle and $\xi$ is the kinetic mixing parameter which needs to be small in order to be compatible with the experimental constraints.
The hat fields are not canonically normalized yet, the subscript $0$ on the mass $M_{X,0}$, $M_{Z,0}$ and on the fields $Z_0$, $X_0$ means that the fields are not in the mass eigenbasis.
We  can make use of a redefinition of the field to normalize the kinetic terms:
\begin{equation}
    \begin{pmatrix}
        X_0\\
        B
    \end{pmatrix}=
    \begin{pmatrix}
        \sqrt{1-\xi^2/c_W^2}&0\\
        -\xi/c_W &1
    \end{pmatrix}
    \begin{pmatrix}
        \hat{X_0}\\
        \hat{B}
    \end{pmatrix}.
\end{equation}
Introducing the following quantities
\begin{equation}
    \eta=\frac{\xi/c_W}{\sqrt{1-\frac{\xi^2}{c_W^2}}}\qquad \qquad\delta^2=\frac{M^2_{X,0}}{M^2_{Z,0}}
\end{equation}
we can diagonalize the gauge boson mass terms by means of a rotation.
The mass eigenvalues are then:

\begin{eqnarray}
   \nonumber M^2_{Z,X}&=&\frac{M^2_{Z,0}}{2}\bigg(1+\delta^2+s^2_W\eta^2 + \\ &\pm& \mathrm{sign} (1-\delta^2) \sqrt{(1+\delta^2+s^2_W\eta)^2-4\delta^2}\bigg).
\end{eqnarray}
At the Z resonance, the primary impact of the $X_{17}$ particle is to alter the mass of the $Z$ boson and its couplings. This influence is described by the so called oblique parameters. To express the oblique parameters in terms of the kinetic mixing parameter, one begins with the kinetic mixing Lagrangian, normalizes the kinetic terms canonically, sets the physical constants to their Standard Model values, and then matches the results to the general form of the mass terms and interactions when oblique corrections are present.
The oblique parameters, in the presence of kinetic mixing, have already been calculated in \cite{W03, W2}. By inserting them into the formula for the W boson mass with oblique corrections, the shift on the $W$ boson mass is given by:
\begin{equation}\label{Wmassshift}
    \Delta M_W=M_W-M_{W,SM}=-\frac{M_{W,SM}s_W^2\xi^2}{2(c_W^2-s_W^2)^2(1-r^2)},
\end{equation}
with $r=\frac{M_Z}{M_X}$. By setting $\Delta M_W\leq 10 \ \mathrm{MeV}$ corresponding to the uncertainty from the latest measurements \cite{W1} we derive, from Eq. \eqref{Wmassshift}, an upper bound on the kinetic mixing parameter: $|\xi|<2.2 \times 10^{-2}$.

\vspace{2mm}

{\it Conclusions} -- Given the experimental hints on the existence of the new light boson $X_{17}$ coming from atomic decays, we have studied its impact on several particle physics phenomena. In particular we have shown that $X_{17}$ may explain some of the most significant anomalies in QED, as well as contribute a correction to the gauge boson masses. In particular, we addressed the anomalies in the magnetic moments of the muon and electron by considering the interaction of these leptons with $X_{17}$. This allowed us to compute an upper bound on the coupling constant between $X_{17}$ and the leptons.
Additionally, we calculated the correction to the Lamb shift in muonic atoms due to the $X_{17}$ vector boson and compared it with experimental data. From this comparison, we derived an upper bound for the coupling constant of $X_{17}$ with neutrons and protons, employing the previously derived estimations on the muon coupling.
Furthermore by considering the kinetic mixing between the $X_{17}$ boson and the $U(1)_Y$ gauge boson, we established an upper limit on the kinetic mixing constant, constrained by the most recent experimental uncertainties.
Considering its role in explaining atomic decays and the results obtained in this work, we conclude that $X_{17}$ represents a reasonable, fruitful and economical extension of the standard model, capable of accounting for or easing some of the current experimental tensions in particle physics.

\section*{Acknowledgements}
We acknowledge partial financial support from MUR and INFN, A.C. also acknowledges the COST Action CA1511 Cosmology
and Astrophysics Network for Theoretical Advances and Training
Actions (CANTATA).

\end{document}